\documentclass{aa}  

\usepackage{graphicx}
\usepackage{txfonts}
\usepackage{color}
\usepackage{stfloats}
\usepackage{multirow}

\newcommand{\tablenotea}[1]{\parbox{18.3cm}{\indent \footnotesize{#1}}}
\newcommand{\acsesc}{ACS Earth Space Chem.}
\newcommand{\cpl}{Chem. Phys. Lett.}
\newcommand{\chemrev}{Chem. Rev.}

\newcommand{\jacs}{J. Am. Chem. Soc.}
\newcommand{\jcsftt}{J. Chem. Soc. Faraday Trans. 2}
\newcommand{\jms}{J. Mol. Spectr.}
\newcommand{\jmst}{J. Mol. Struct.}

\newcommand{\jpca}{J. Phys. Chem. A}

\newcommand{\pccp}{Phys. Chem. Chem. Phys.}

\newcommand{\pnas}{PNAS}

\newcommand{\tca}{Theor. Chem. Acc.}

\begin{document}

\title{O-bearing complex organic molecules at the cyanopolyyne peak of \mbox{TMC-1}: detection of C$_2$H$_3$CHO, C$_2$H$_3$OH, HCOOCH$_3$, and CH$_3$OCH$_3$\thanks{Based on observations carried out with the Yebes 40m telescope (projects 19A003, 20A014, and 20D023). The 40m radiotelescope at Yebes Observatory is operated by the Spanish Geographic Institute (IGN, Ministerio de Transportes, Movilidad y Agenda Urbana).}}

\titlerunning{O-bearing complex organic molecules in \mbox{TMC-1}}
\authorrunning{Ag\'undez et al.}

\author{M.~Ag\'undez\inst{1}, N.~Marcelino\inst{1}, B.~Tercero\inst{2,3}, C.~Cabezas\inst{1}, P.~de~Vicente\inst{3}, \and J.~Cernicharo\inst{1}}

\institute{
Instituto de F\'isica Fundamental, CSIC, Calle Serrano 123, E-28006 Madrid, Spain\\ \email{marcelino.agundez@csic.es, jose.cernicharo@csic.es} \and
Observatorio Astron\'omico Nacional, IGN, Calle Alfonso XII 3, E-28014 Madrid, Spain \and
Observatorio de Yebes, IGN, Cerro de la Palera s/n, E-19141 Yebes, Guadalajara, Spain
}

\date{Received; accepted}

 
\abstract
{We report the detection of the oxygen-bearing complex organic molecules propenal (C$_2$H$_3$CHO), vinyl alcohol (C$_2$H$_3$OH), methyl formate (HCOOCH$_3$), and dimethyl ether (CH$_3$OCH$_3$) toward the cyanopolyyne peak of the starless core \mbox{TMC-1}. These molecules are detected through several emission lines in a deep Q-band line survey of \mbox{TMC-1} carried out with the Yebes 40m telescope. These observations reveal that the cyanopolyyne peak of \mbox{TMC-1}, which is the prototype of cold dark cloud rich in carbon chains, contains also O-bearing complex organic molecules like HCOOCH$_3$ and CH$_3$OCH$_3$, which have been previously seen in a handful of cold interstellar clouds. In addition, this is the first secure detection of C$_2$H$_3$OH in space and the first time that C$_2$H$_3$CHO and C$_2$H$_3$OH are detected in a cold environment, adding new pieces in the puzzle of complex organic molecules in cold sources. We derive column densities of (2.2\,$\pm$\,0.3)\,$\times$\,10$^{11}$\,cm$^{-2}$, (2.5\,$\pm$\,0.5)\,$\times$\,10$^{12}$\,cm$^{-2}$, (1.1\,$\pm$\,0.2)\,$\times$\,10$^{12}$\,cm$^{-2}$, and (2.5\,$\pm$\,0.7)\,$\times$\,10$^{12}$\,cm$^{-2}$ for C$_2$H$_3$CHO, C$_2$H$_3$OH, HCOOCH$_3$, and CH$_3$OCH$_3$, respectively. Interestingly, C$_2$H$_3$OH has an abundance similar to that of its well known isomer acetaldehyde (CH$_3$CHO), with C$_2$H$_3$OH/CH$_3$CHO\,$\sim$\,1 at the cyanopolyyne peak. We discuss potential formation routes to these molecules and recognize that further experimental, theoretical, and astronomical studies are needed to elucidate the true mechanism of formation of these O-bearing complex organic molecules in cold interstellar sources.}


\maketitle

\section{Introduction}

Complex organic molecules (COMs) like methyl formate (HCOOCH$_3$) and dimethyl ether (CH$_3$OCH$_3$) have been traditionally observed in the warm gas around protostars, the so-called hot cores and corinos, where they are thought to form upon thermal desorption of ice mantles on grains \citep{Herbst2009}. In the last decade, these molecules have been observed as well in a few cold sources, like the dense cores B1-b \citep{Oberg2010,Cernicharo2012} and L483 \citep{Agundez2019}, the dark cloud Barnard\,5 \citep{Taquet2017}, the pre-stellar cores L1689B \citep{Bacmann2012} and L1544 \citep{Jimenez-Serra2016}, and the starless core \mbox{TMC-1} \citep{Soma2018}. The low temperatures in these environments inhibit thermal desorption, and it is still an active subject of debate how are these molecules formed, whether in the gas phase or on grain surfaces followed by some non-thermal desorption process \citep{Vasyunin2013,Ruaud2015,Balucani2015,Chang2016,Vasyunin2017,Shingledecker2018,Jin2020}.

The cyanopolyyne peak of \mbox{TMC-1}, \mbox{TMC-1(CP)}, is characterized by a carbon-rich chemistry, with high abundances of carbon chains and a poor content of O-bearing COMs (e.g., \citealt{Agundez2013}). Here we report the detection of four O-bearing COMs toward \mbox{TMC-1(CP)}. Propenal (C$_2$H$_3$CHO) has been reported previously toward massive star-forming regions in the Galactic center \citep{Hollis2004,Requena-Torres2008} and in the hot corino IRAS\,16293-2422\,B \citep{Manigand2021}. Vinyl alcohol (C$_2$H$_3$OH) has been only seen toward Sagittarius\,B2(N), where the high spectral density complicates the identification \citep{Turner2001}. Therefore, this is the first clear detection of C$_2$H$_3$OH in space and the first time that C$_2$H$_3$CHO and C$_2$H$_3$OH are detected in a cold environment. We also report the detection of HCOOCH$_3$ and CH$_3$OCH$_3$, recently detected (the latter tentatively) toward the methanol peak of \mbox{TMC-1} \citep{Soma2018} but not toward TMC-1(CP).

\section{Astronomical observations}

The data presented here belong to a Q-band line survey of \mbox{TMC-1(CP)}, $\alpha_{J2000}$\,=\,$4^{\rm h}$\,$41^{\rm  m}$\,$41.9^{\rm s}$ and $\delta_{J2000}$\,=\,$+25^\circ$\,$41'$\,$27.0''$, performed with the Yebes 40m telescope. The cryogenic receiver for the Q band, built within the Nanocosmos project\,\footnote{\texttt{https://nanocosmos.iff.csic.es}} and which covers the 31.0-50.4\,GHz frequency range with horizontal and vertical polarizations, was used connected to FFTS spectrometers, which cover a bandwidth of $8\times2.5$\,GHz in each polarization with a spectral resolution of 38.15\,kHz. The system is described in \cite{Tercero2021}. The half power beam width (HPBW) of the Yebes 40m telescope ranges from 36.4\,$''$ to 54.4\,$''$ in the Q band. The intensity scale is antenna temperature, $T_A^*$, for which we estimate an uncertainty of 10\,\%, which can be converted to main beam brightness temperature, $T_{\rm mb}$, by dividing by $B_{\rm eff}$/$F_{\rm eff}$ (see Table~\ref{table:lines}). The line survey was carried out during several observing runs and various results have been already published. Data taken during November 2019 and February 2020 allowed to detect the negative ions C$_3$N$^-$ and C$_5$N$^-$ \citep{Cernicharo2020a}, and to discover HC$_4$NC \citep{Cernicharo2020b}, HC$_3$O$^+$ \citep{Cernicharo2020c}, and HC$_5$NH$^+$ \citep{Marcelino2020}. A further observing run, carried out in October 2020, resulted in the detection of HDCCN \citep{Cabezas2021}, HC$_3$S$^+$ \citep{Cernicharo2021a}, CH$_3$CO$^+$ \citep{Cernicharo2021b}, and various C$_4$H$_3$N isomers \citep{Marcelino2021}. Additional observations were taken in December 2020 and January 2021, which led to the discovery of vinyl acetylene (CH$_2$CHCCH; \citealt{Cernicharo2021c}), allenyl acetylene (CH$_2$CCHCCH; \citealt{Cernicharo2021d}), and propargyl (CH$_2$CCH; \citealt{Agundez2021}), and a final run was carried out in March 2021. All observations were carried out using the frequency switching technique, with a frequency throw of 10\,MHz during the two first observing runs and 8\,MHz in the later ones. All data were reduced with the program {\small CLASS} of {\small GILDAS} \citep{Pety2005}\,\footnote{\texttt{http://www.iram.fr/IRAMFR/GILDAS}}.

\begin{table*}
\small
\caption{Observed line parameters of the target O-bearing COMs of this study in \mbox{TMC-1}.}
\label{table:lines}
\centering
\begin{tabular}{llrlccccrr}
\hline \hline
\multicolumn{1}{l}{Molecule} & \multicolumn{1}{l}{Transition} & \multicolumn{1}{c}{$E_{\rm up}$} & \multicolumn{1}{c}{$\nu_{\rm calc}$} & \multicolumn{1}{c}{$T_A^*$ peak} & \multicolumn{1}{c}{$\Delta v$\,$^a$}      & \multicolumn{1}{c}{$V_{\rm LSR}$}      & \multicolumn{1}{c}{$\int T_A^* dv$} & \multicolumn{1}{c}{S/N\,$^b$} & \multicolumn{1}{c}{\textcolor{magenta}{$B_{\rm eff}$/$F_{\rm eff}$\,$^c$}} \\
 & & \multicolumn{1}{c}{(K)}        & \multicolumn{1}{c}{(MHz)}        & \multicolumn{1}{c}{(mK)}                   & \multicolumn{1}{c}{(km s$^{-1}$)}  & \multicolumn{1}{c}{(km s$^{-1}$)}  & \multicolumn{1}{c}{(mK km s$^{-1}$)} & \multicolumn{1}{c}{($\sigma$)} & \\
\hline
\multirow{6}{*}{$trans$-C$_2$H$_3$CHO} & 4$_{1,4}$-3$_{1,3}$ & 6.2 & 34768.987 & $3.06 \pm 0.31$ & $0.62 \pm 0.08$ & $5.77 \pm 0.03$ & $2.03 \pm 0.21$ & 14.5 & 0.603 \\ 
                                                       & 4$_{0,4}$-3$_{0,3}$ & 4.3 & 35578.136 & $3.77 \pm 0.33$ & $0.81 \pm 0.08$ & $5.79 \pm 0.03$ & $3.27 \pm 0.27$ & 19.4 & 0.597 \\ 
                                                       & 4$_{1,3}$-3$_{1,2}$ & 6.4 & 36435.990 & $3.17 \pm 0.29$ & $0.85 \pm 0.09$ & $5.73 \pm 0.04$ & $2.87 \pm 0.25$ & 19.2 & 0.590 \\ 
                                                       & 5$_{1,5}$-4$_{1,4}$ & 8.3 & 43455.469 & $1.81 \pm 0.49$ & $0.72 \pm 0.27$ & $5.78 \pm 0.08$ & $1.38 \pm 0.37$ & 6.5 & 0.530 \\ 
                                                       & 5$_{0,5}$-4$_{0,4}$ & 6.4 & 44449.749 & $3.67 \pm 0.54$ & $0.54 \pm 0.08$ & $5.80 \pm 0.04$ & $2.11 \pm 0.31$ & 10.5 & 0.521 \\ 
                                                       & 5$_{1,4}$-4$_{1,3}$ & 8.6 & 45538.994 & $4.72 \pm 0.66$ & $0.69 \pm 0.09$ & $5.80 \pm 0.04$ & $3.45 \pm 0.43$ & 12.6 & 0.511 \\ 
\hline
\multirow{4}{*}{$syn$-C$_2$H$_3$OH} & 4$_{0,4}$-3$_{1,3}$ & 9.3 & 32449.221 & $1.17 \pm 0.32$ & $1.07 \pm 0.21$ & $5.61 \pm 0.11$ & $1.34 \pm 0.27$ & 6.8 & 0.622 \\ 
                                                               & 2$_{1,2}$-1$_{1,1}$ & 5.1 & 37459.184\,$^d$ & -- & -- & -- & -- & -- & 0.581 \\ 
                                                               & 2$_{0,2}$-1$_{0,1}$ & 2.8 & 39016.387 & $1.90 \pm 0.39$ & $0.89 \pm 0.16$ & $5.81 \pm 0.07$ & $1.79 \pm 0.29$ & 9.0 & 0.568 \\ 
                                                               & 2$_{1,1}$-1$_{1,0}$ & 5.3 & 40650.606 & $1.71 \pm 0.41$ & $0.65 \pm 0.19$ & $5.88 \pm 0.08$ & $1.18 \pm 0.28$ & 6.7 & 0.554 \\ 
\hline
\multirow{10}{*}{HCOOCH$_3$} & 3$_{1,3}$-2$_{1,2}$ $E$ &4.0 &  34156.889 & $1.92 \pm 0.29$ & $0.62 \pm 0.12$ & $5.93 \pm 0.06$ & $1.27 \pm 0.23$ & 9.6 & 0.608 \\ 
                                                    & 3$_{1,3}$-2$_{1,2}$ $A$ & 3.9 & 34158.061 & $1.86 \pm 0.29$ & $0.80 \pm 0.16$ & $5.84 \pm 0.07$ & $1.59 \pm 0.28$ & 10.6 & 0.608 \\ 
                                                    & 3$_{0,3}$-2$_{0,2}$ $E$ & 3.5 & 36102.227\,$^e$ & -- & -- & -- & -- & -- & 0.593 \\ 
                                                    & 3$_{0,3}$-2$_{0,2}$ $A$ & 3.5 & 36104.775 & $2.61 \pm 0.29$ & $0.74 \pm 0.09$ & $5.80 \pm 0.04$ & $2.05 \pm 0.22$ & 14.6 & 0.592 \\ 
                                                    & 3$_{1,2}$-2$_{1,1}$ $E$ & 4.4 & 38976.085 & $1.81 \pm 0.35$ & $0.66 \pm 0.14$ & $5.93 \pm 0.07$ & $1.28 \pm 0.25$ & 8.3 & 0.568 \\ 
                                                    & 3$_{1,2}$-2$_{1,1}$ $A$ & 4.4 & 38980.803 & $2.45 \pm 0.35$ & $0.57 \pm 0.09$ & $5.92 \pm 0.04$ & $1.49 \pm 0.21$ & 10.4 & 0.568 \\ 
                                                    & 4$_{1,4}$-3$_{1,3}$ $E$ & 6.1 & 45395.802 & $2.82 \pm 0.57$ & $0.63 \pm 0.18$ & $6.00 \pm 0.07$ & $1.89 \pm 0.40$ & 8.3 & 0.512 \\ 
                                                    & 4$_{1,4}$-3$_{1,3}$ $A$ & 6.1 & 45397.360 & $2.39 \pm 0.57$ & $0.42 \pm 0.19$ & $5.91 \pm 0.10$ & $1.07 \pm 0.32$ & 5.8 & 0.512 \\ 
                                                    & 4$_{0,4}$-3$_{0,3}$ $E$ & 5.8 & 47534.116 & $3.19 \pm 0.75$ & $0.59 \pm 0.12$ & $5.92 \pm 0.07$ & $2.01 \pm 0.44$ & 7.1 & 0.493 \\ 
                                                    & 4$_{0,4}$-3$_{0,3}$ $A$ & 5.8 & 47536.905 & $2.76 \pm 0.75$ & $0.98 \pm 0.23$ & $5.99 \pm 0.10$ & $2.88 \pm 0.60$ & 7.9 & 0.493 \\ 
\hline
\multirow{10}{*}{CH$_3$OCH$_3$} & 2$_{1,1}$-2$_{0,2}$ $AE + EA$ & 4.2 & 31105.223 & $1.35 \pm 0.39$ & $0.92 \pm 0.25$ & $5.72 \pm 0.13$ & $1.31 \pm 0.36$ & 5.8 & 0.632 \\ 
                                                         & 2$_{1,1}$-2$_{0,2}$ $EE$ & 4.2 & 31106.150 & $1.52 \pm 0.39$ & $0.54 \pm 0.34$ & $5.94 \pm 0.08$ & $0.87 \pm 0.29$ & 5.0 & 0.632 \\ 
                                                         & 3$_{1,2}$-3$_{0,3}$ $AE + EA$ & 7.0 & 32977.276 & $1.37 \pm 0.25$ & $0.63 \pm 0.13$ & $5.84 \pm 0.07$ & $0.91 \pm 0.19$ & 7.8 & 0.618 \\ 
                                                         & 3$_{1,2}$-3$_{0,3}$ $EE$ & 7.0 & 32978.232 & $1.37 \pm 0.25$ & $0.95 \pm 0.23$ & $5.84 \pm 0.09$ & $1.38 \pm 0.27$ & 9.6 & 0.618 \\ 
                                                         & 3$_{1,2}$-3$_{0,3}$ $AA$ & 7.0 & 32979.187 & $1.06 \pm 0.25$ & $0.85 \pm 0.32$ & $5.99 \pm 0.12$ & $0.96 \pm 0.29$ & 7.1 & 0.618 \\ 
                                                         & 4$_{1,3}$-4$_{0,4}$ $EE$ & 10.8 & 35593.408 & $1.22 \pm 0.27$ & $0.56 \pm 0.13$ & $5.82 \pm 0.07$ & $0.73 \pm 0.16$ & 6.4 & 0.597 \\ 
                                                         & 1$_{1,1}$-0$_{0,0}$ $EE$ & 2.3 & 47674.967 & $3.31 \pm 0.62$ & $0.59 \pm 0.12$ & $6.02 \pm 0.06$ & $2.08 \pm 0.38$ & 8.9 & 0.492 \\ 
\hline
\end{tabular}
\tablenotea{\\
The line parameters $T_A^*$ peak, $\Delta v$, $V_{\rm LSR}$, and $\int T_A^* dv$ and the associated errors are derived from a Gaussian fit to each line profile. $^a$\,$\Delta v$ is the full width at half maximum (FWHM). $^b$\,Signal-to-noise ratio is computed as S/N = $\int T_A^* dv$ / [rms $\times$ $\sqrt{\Delta v \times \delta \nu (c / \nu_{\rm calc})}$], where $c$ is the speed of light, $\delta \nu$ is the spectral resolution (0.03815 MHz), the rms is given in the uncertainty of $T_A^*$ peak, and the rest of parameters are given in the table. $^c$\,$B_{\rm eff}$ is given by the Ruze formula $B_{\rm eff}$\,=\,0.738\,$\exp{[-(\nu/72.2)^2]}$, where $\nu$ is the frequency in GHz, and $F_{\rm eff}$\,=\,0.97. $^d$\,Line overlaps with a hyperfine component of CH$_2$CCH (see \citealt{Agundez2021}). $^e$\,Line detected but not fitted because it overlaps with a negative frequency-switching artifact.
}
\end{table*}

\section{Molecular spectroscopy} \label{sec:spectroscopy}

\begin{figure}
\centering
\includegraphics[angle=0,width=0.879\columnwidth]{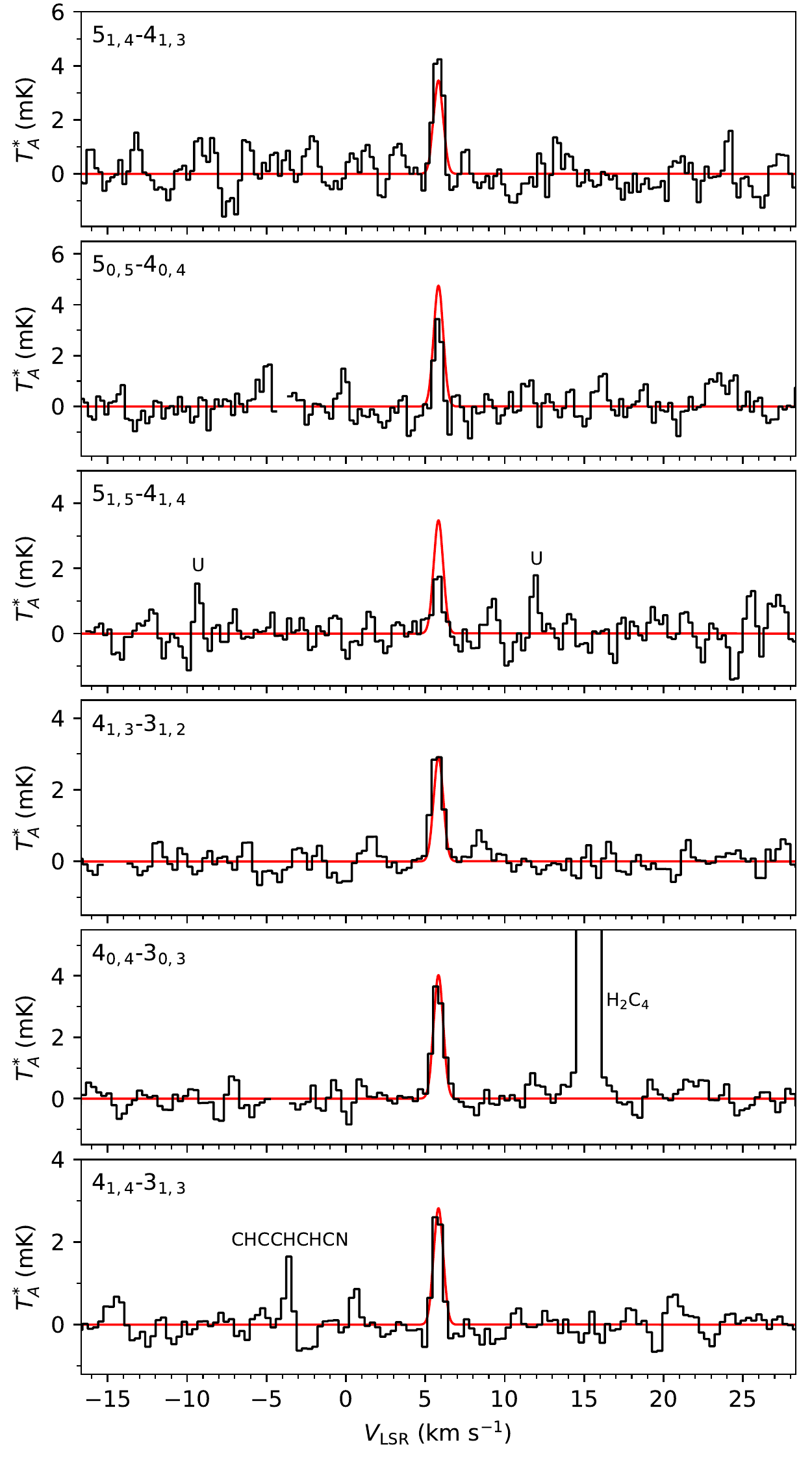}
\caption{Lines of C$_2$H$_3$CHO observed in \mbox{TMC-1} (see parameters in Table~\ref{table:lines}). In red computed synthetic spectra for $N$\,=\,2.2\,$\times$\,10$^{11}$\,cm$^{-2}$, $T_{\rm rot}$\,=\,7.5\,K, FWHM\,=\,0.71 km s$^{-1}$, and $\theta_s$\,=\,80\,$''$.} \label{fig:c2h3cho}
\end{figure}

\begin{figure}
\centering
\includegraphics[angle=0,width=0.879\columnwidth]{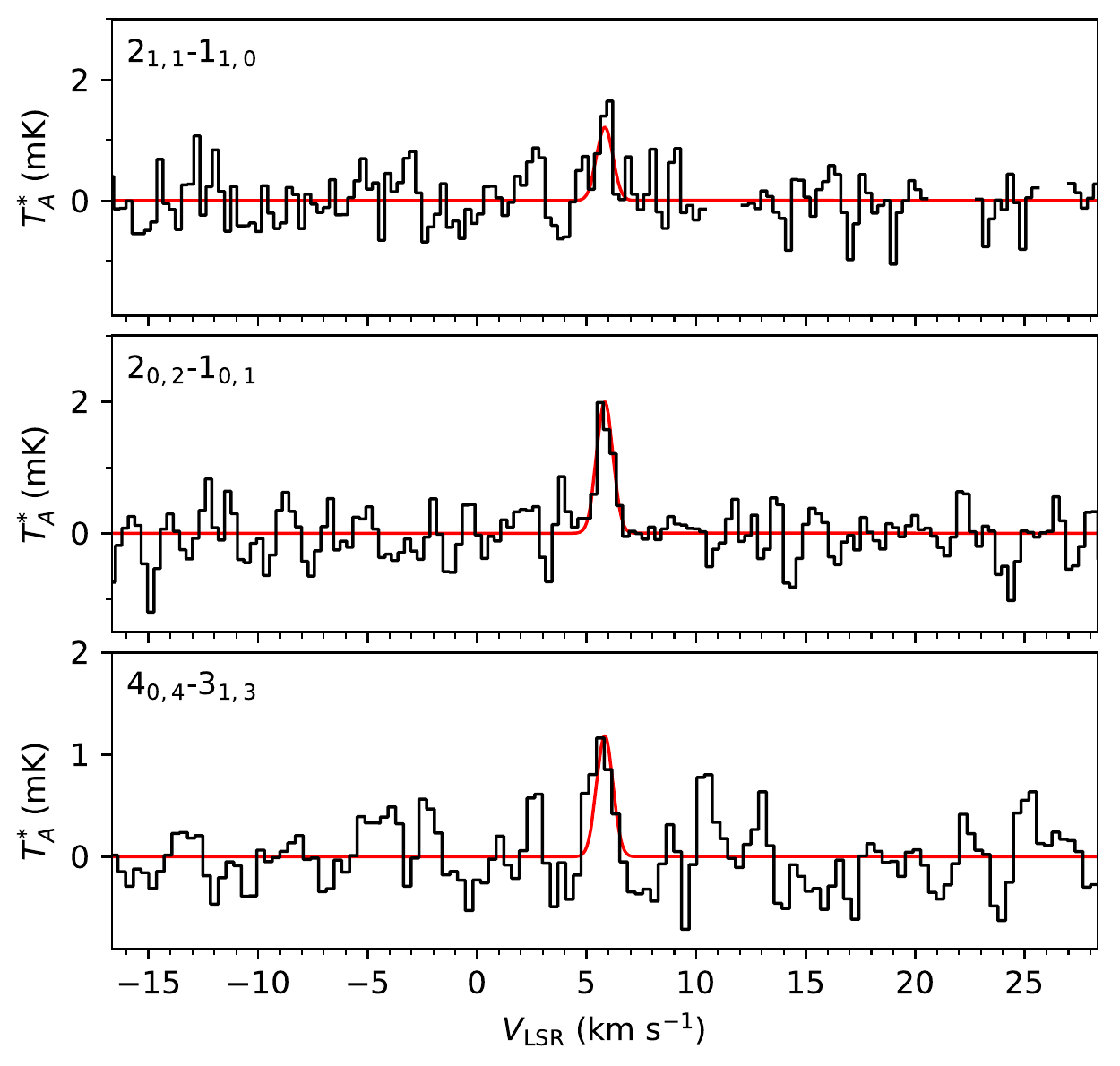}
\caption{Lines of C$_2$H$_3$OH observed in \mbox{TMC-1} (see parameters and note on line 2$_{1,1}$-1$_{1,0}$ in Table~\ref{table:lines}). In red computed synthetic spectra for $N$\,=\,2.5\,$\times$\,10$^{12}$\,cm$^{-2}$, $T_{\rm rot}$\,=\,10\,K, FWHM\,=\,0.87 km s$^{-1}$, and $\theta_s$\,=\,80\,$''$.} \label{fig:c2h3oh}
\end{figure}

\begin{figure}
\centering
\includegraphics[angle=0,width=0.836\columnwidth]{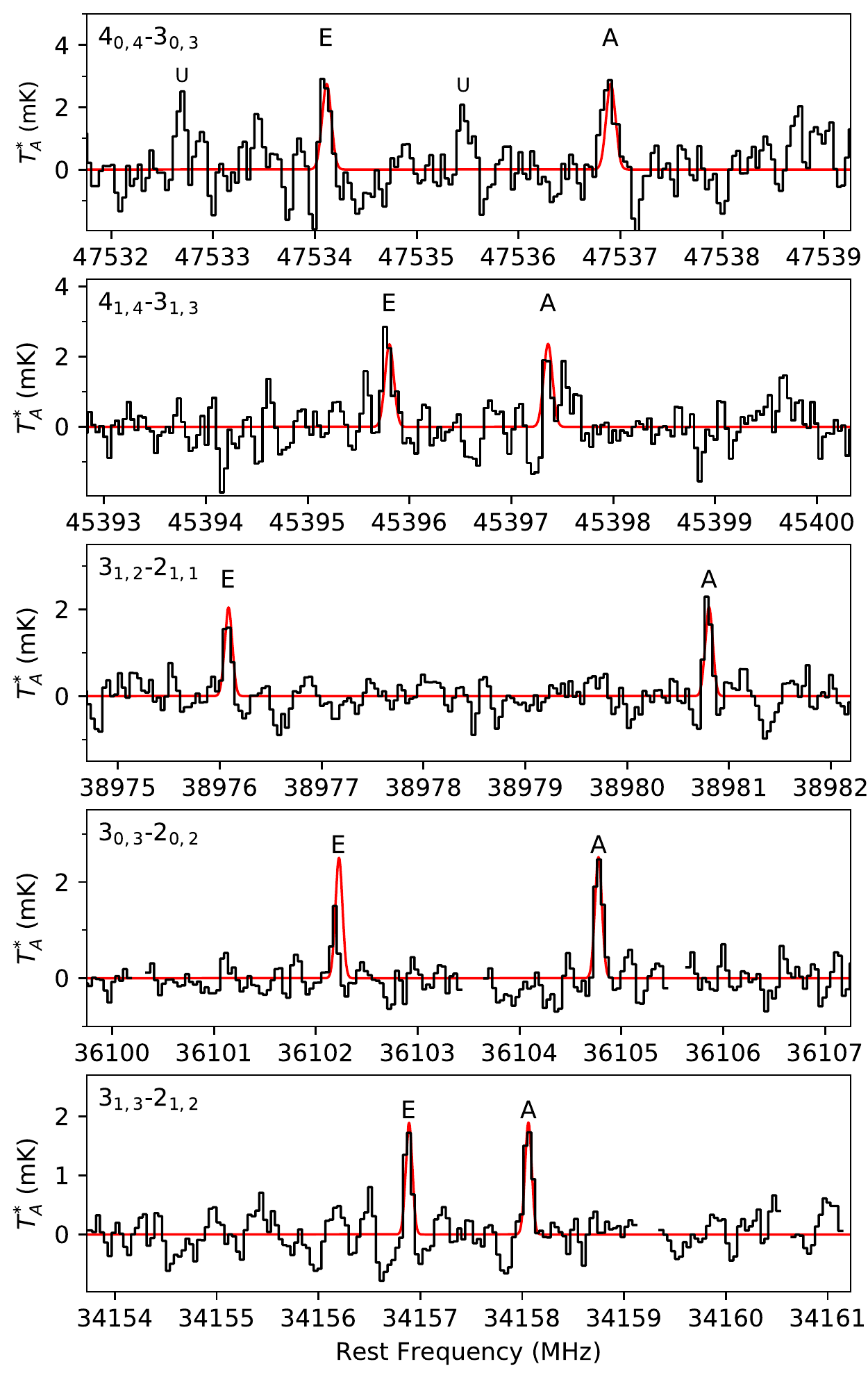}
\caption{Lines of HCOOCH$_3$ observed in \mbox{TMC-1} (see parameters and note on line 3$_{0,3}$-2$_{0,2}$ $E$ in Table~\ref{table:lines}). In red computed synthetic spectra for a $N$\,=\,1.1\,$\times$\,10$^{12}$\,cm$^{-2}$, $T_{\rm rot}$\,=\,5\,K, FWHM\,=\,0.67 km s$^{-1}$, and $\theta_s$\,=\,80\,$''$.} \label{fig:hcooch3}
\end{figure}

\begin{figure}
\centering
\includegraphics[angle=0,width=0.836\columnwidth]{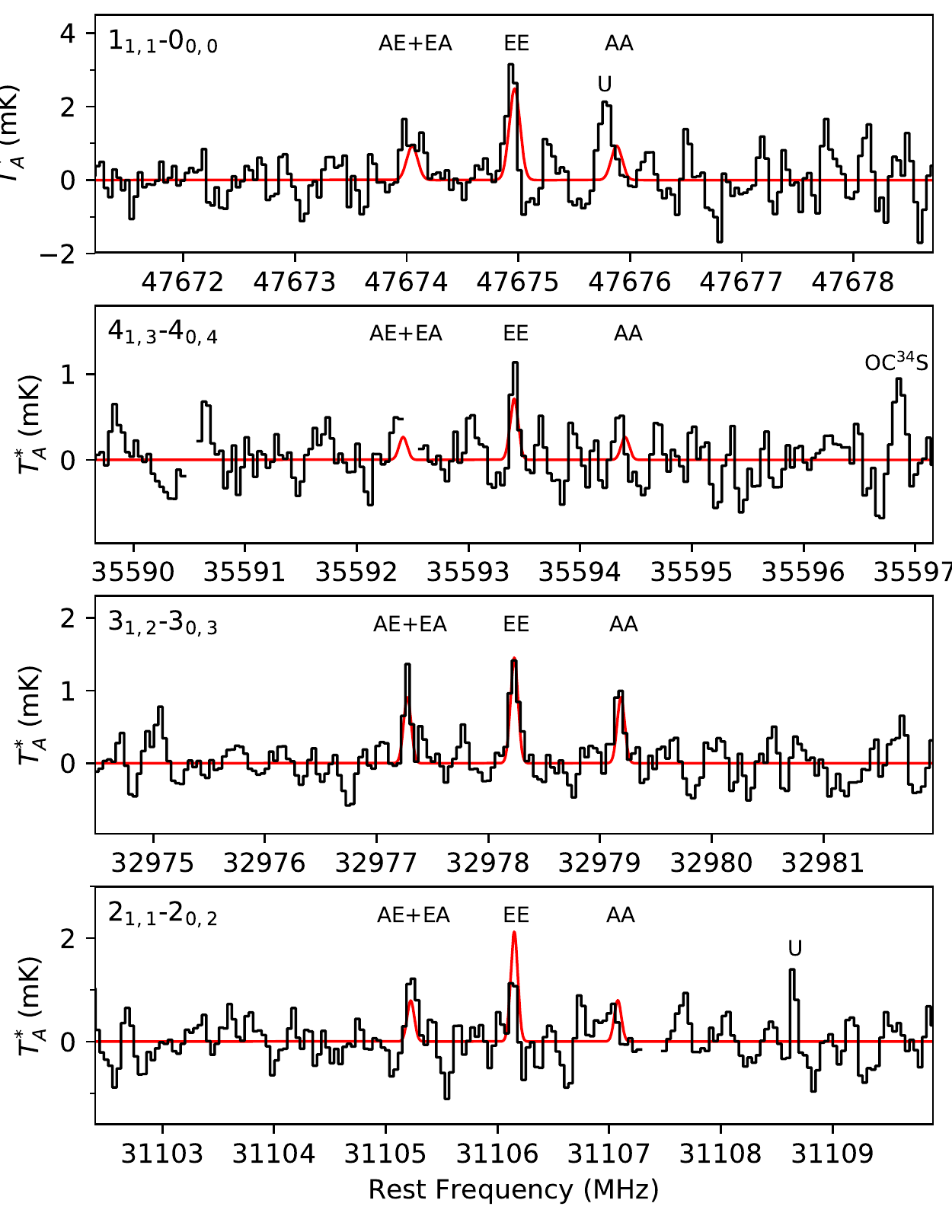}
\caption{Lines of CH$_3$OCH$_3$ observed in \mbox{TMC-1} (see parameters in Table~\ref{table:lines}). In red computed synthetic spectra for $N$\,=\,2.5\,$\times$\,10$^{12}$\,cm$^{-2}$, $T_{\rm rot}$\,=\,3.6\,K, FWHM\,=\,0.72 km s$^{-1}$, and $\theta_s$\,=\,80\,$''$.} \label{fig:ch3och3}
\end{figure}

C$_2$H$_3$CHO has two conformers. The most stable, and the only one reported in space \citep{Hollis2004,Requena-Torres2008,Manigand2021}, is the $trans$ form, which is the one reported here as well. Level energies and transition frequencies were obtained from the rotational constants derived by \cite{Daly2015}. The dipole moment along the $a$ axis (all transitions observed here are of $a$-type) is 3.052\,D \citep{Blom1984}.

C$_2$H$_3$OH has also two conformers, named $syn$ and $anti$. \cite{Turner2001} assigned various emission features to the two conformers in the crowded spectra of Sgr\,B2(N). Here we report an unambiguous detection of the $syn$ form, which is the most stable one, in a colder source that is much less affected by line confusion. Level energies and transition frequencies were computed from the rotational constants given by \cite{Melosso2019}. The components of the dipole moment along the $a$ and $b$ axes are 0.616\,D and 0.807\,D, respectively \citep{Saito1976}. Both $a$- and $b$-type transitions are observed here.

HCOOCH$_3$ is a well known interstellar molecule in which the internal rotation, or torsion, of the methyl group makes each rotational level to split into $A$ and $E$ substates, with statistical weights A:E\,=\,1:1. We used the spectroscopy from a fit to the lines measured by \cite{Ogata2004} implemented in {\small MADEX} \citep{Cernicharo2012}. Here all observed transitions are of $a$-type and thus we adopted $\mu_a$\,=\,1.63\,D \citep{Curl1959}.

CH$_3$OCH$_3$ is an asymmetric rotor in which the large amplitude internal motion of the two equivalent methyl groups leads to level splitting into $AA$, $EE$, $EA$, and $AE$ substates, with non-trivial statistical weights \citep{Endres2009}. We adopted the spectroscopy from the {\small CDMS} catalog \citep{Muller2005}\,\footnote{\texttt{https://cdms.astro.uni-koeln.de/}}. The geometry of the molecule makes it to have non-zero dipole moment only along the $b$ axis, with a measured value of 1.302\,D \citep{Blukis1963}.

\section{Results}

We detected various lines for each of the O-bearing COMs target of this study (see Table~\ref{table:lines}), with signal-to-noise ratios (S/N) well above 3\,$\sigma$. The position of the lines is consistent with the calculated frequencies based on laboratory data (see Sec.~\ref{sec:spectroscopy}) and the systemic velocity of the source, $V_{\rm LSR}$\,=\,5.83 km s$^{-1}$ \citep{Cernicharo2020b}. Moreover, the relative intensities of the lines are those expected for rotational temperatures in the range 3-10\,K, which are typical of \mbox{TMC-1} \citep{Gratier2016}, and there are no missing lines. We thus consider that the detection of the four O-bearing COMs in \mbox{TMC-1} is secure. Hereafter we discuss the particularities of each molecule.

The detection of the $trans$ form of C$_2$H$_3$CHO is very solid since the six observed lines are detected with S/N between 6.5\,$\sigma$ and 19.4\,$\sigma$ and the $V_{\rm LSR}$ of the lines are fully consistent with the systemic velocity of \mbox{TMC-1} (see Table~\ref{table:lines} and Fig.~\ref{fig:c2h3cho}). The rotational temperature ($T_{\rm rot}$) of C$_2$H$_3$CHO is not very precisely determined, 7.5\,$\pm$\,3.5\,K from a rotation diagram, but synthetic spectra computed under thermodynamic equilibrium indicate that values in the range 5-10\,K are consistent with the relative intensities observed. We thus adopted $T_{\rm rot}$\,=\,7.5\,K to compute the synthetic spectra and derive the column density. The arithmetic mean of the observed C$_2$H$_3$CHO linewidths, 0.71 km s$^{-1}$, is adopted when computing the synthetic spectra (see Table~\ref{table:lines}). We also assumed a circular emission distribution with a diameter $\theta_s$\,=\,80$''$, as observed for various hydrocarbons in \mbox{TMC-1} \citep{Fosse2001}. For the other three molecules we followed the same convention, adopting as linewidth the average of the observed values and assuming the same emission distribution. The column density derived for C$_2$H$_3$CHO is (2.2\,$\pm$\,0.3)\,$\times$\,10$^{11}$\,cm$^{-2}$.

In the case of C$_2$H$_3$OH (see Table~\ref{table:lines} and Fig.~\ref{fig:c2h3oh}), three lines are clearly detected, with S/N of 6.7\,$\sigma$ and 9.0\,$\sigma$. The frequency of the $2_{1,2}$-1$_{1,1}$ transition, 37459.184 MHz, coincides with a hyperfine component of CH$_2$CCH, recently reported in \mbox{TMC-1} \citep{Agundez2021}. We predict $T_A^*$ = 1.1\,mK for the $2_{1,2}$-1$_{1,1}$ transition, which indicates that the observed line (see \citealt{Agundez2021}) has contributions from both CH$_2$CCH and C$_2$H$_3$OH. The high detection significance of three lines, the overlap of a fourth line with CH$_2$CCH, and the fact that there are no missing lines, makes us to consider the detection of C$_2$H$_3$OH secure. The rotational temperature is not well constrained for C$_2$H$_3$OH, although the observed relative intensities indicate that it must be in the high range of the values typically observed in \mbox{TMC-1}. We thus adopted $T_{\rm rot}$\,=\,10\,K and a linewidth of 0.87 km s$^{-1}$ to compute the synthetic spectra. We derive a column density of (2.5\,$\pm$\,0.5)\,$\times$\,10$^{12}$\,cm$^{-2}$ for C$_2$H$_3$OH.

We detected five $A$/$E$ doublets of HCOOCH$_3$ with S/N in the range 5.8-14.6\,$\sigma$ (see Table~\ref{table:lines} and Fig.~\ref{fig:hcooch3}). The only line affected by a problem is the 3$_{0,3}$-2$_{0,2}$ transition of the $E$ substate, which accidentally lies close to a negative frequency-switching artifact, making it to appear less intense and slightly shifted from the correct position. We thus did not fit this line. The rotational temperature derived is 5.1\,$\pm$\,2.5\,K, and we thus adopted $T_{\rm rot}$\,=\,5 K, and a linewidth of 0.67 km s$^{-1}$, to compute the synthetic spectra. The total column density obtained for HCOOCH$_3$, including both $A$ and $E$ substates, is (1.1\,$\pm$\,0.2)\,$\times$\,10$^{12}$\,cm$^{-2}$.

For CH$_3$OCH$_3$, we observed four triplets with the characteristic structure of an intense component corresponding to the $EE$ substate lying between two equally intense components corresponding to the $AE$+$EA$ and $AA$ substates. The $EE$ component of the four triplets is detected with good confidence levels, between 5.0\,$\sigma$ and 9.6\,$\sigma$ (see Table~\ref{table:lines} and Fig.~\ref{fig:ch3och3}). The weaker components corresponding to the $AE$+$EA$ and $AA$ substates are sometimes found to lie within the noise, although the computed synthetic spectra is consistent with this fact. We consider that the detection of CH$_3$OCH$_3$ is secure. From a rotation diagram we derive a low rotational temperature of 3.6\,$\pm$\,0.6\,K, which is well constrained by the availability of transition covering upper level energies from 2.3\,K to 10.8\,K. We thus adopted $T_{\rm rot}$\,=\,3.6\,K and a linewidth of 0.72 km s$^{-1}$ to compute the synthetic spectra, which implies a total column density, including the four substates, of (2.5\,$\pm$\,0.7)\,$\times$\,10$^{12}$\,cm$^{-2}$ for CH$_3$OCH$_3$.

The variation in the column densities due to the uncertainty in $T_{\rm rot}$ is small, $\sim$\,15\,\%, for C$_2$H$_3$CHO and C$_2$H$_3$OH, and higher, a factor of two, for HCOOCH$_3$ and CH$_3$OCH$_3$.

\section{Discussion}

The abundances derived for C$_2$H$_3$CHO, C$_2$H$_3$OH, HCOOCH$_3$, and CH$_3$OCH$_3$ are 2.2\,$\times$\,10$^{-11}$, 2.5\,$\times$\,10$^{-10}$, 1.1\,$\times$\,10$^{-10}$, and 2.5\,$\times$\,10$^{-10}$, respectively, relative to H$_2$, if we adopt a column density of H$_2$ of 10$^{22}$ cm$^{-2}$ \citep{Cernicharo1987}. Now, how are these four O-bearing COMs formed in \mbox{TMC-1}?.

C$_2$H$_3$CHO and C$_2$H$_3$OH could be formed by gas-phase neutral-neutral reactions between reactive radicals like OH, CH, or C$_2$H and abundant closed-shell molecules. However, among the potential sources of C$_2$H$_3$OH, the reactions OH + C$_2$H$_4$ and OH + CH$_2$CHCH$_3$ seem to have activation barriers \citep{Zhu2005,Zador2009} and the reaction CH + CH$_3$OH seems to yield H$_2$CO and CH$_3$ as products \citep{Zhang2002}. In the case of C$_2$H$_3$CHO, a potential formation reaction is OH + allene, but the main products are H$_2$CCO + CH$_3$ \citep{Daranlot2012}. Two more promising routes to C$_2$H$_3$CHO are the reactions CH + CH$_3$CHO, which has been found to produce C$_2$H$_3$CHO \citep{Goulay2012}, and C$_2$H + CH$_3$OH, which to our knowledge has not been studied experimentally or theoretically.

C$_2$H$_3$CHO and C$_2$H$_3$OH are not specifically considered in the chemical networks {\small UMIST RATE12} \citep{McElroy2013} or {\small KIDA uva.kida.2014} \citep{Wakelam2015}, but acetaldehyde (CH$_3$CHO), which is an isomer of C$_2$H$_3$OH, is included. Since it often happens that astrochemical databases do not distinguish between different isomers because information is not available, it is conceivable that some of the reactions that are considered to produce CH$_3$CHO could also form C$_2$H$_3$OH. According to a standard pseudo-time-dependent gas-phase chemical model of a cold dark cloud (e.g., \citealt{Agundez2013}), there are two main reactions of formation of CH$_3$CHO. The first is O + C$_2$H$_5$, for which formation of C$_2$H$_3$OH does not seem to be an important channel, according to experiments \citep{Slagle1988} and theory \citep{Jung2011,Vazart2020}. The second is the dissociative recombination of CH$_3$CHOH$^+$ with electrons, in which case experiments show that the CCO backbone is preserved with a branching ratio of 23\,\% \citep{Hamberg2010}, so that it is possible that both CH$_3$CHO and C$_2$H$_3$OH are formed. A different route to CH$_3$CHO starting from abundant ethanol (C$_2$H$_5$OH) has been proposed \citep{Skouteris2018,Vazart2020}, but in L483, the only cold environment where C$_2$H$_5$OH has been detected, its abundance is twice smaller than that of CH$_3$CHO \citep{Agundez2019}. The column density of CH$_3$CHO at \mbox{TMC-1(CP)} is (2.7-3.5)\,$\times$\,10$^{12}$ cm$^{-2}$ \citep{Gratier2016,Cernicharo2020c}, which implies a C$_2$H$_3$OH/CH$_3$CHO ratio of $\sim$\,1. Therefore, if the two isomers are formed by the same reaction, then the branching ratios should be similar. The reaction CH$_3$$^+$ + H$_2$CO produces CH$_4$ + HCO$^+$ \citep{Smith1978}, and thus it is unlikely to form C$_2$H$_3$OH, as suggested by \cite{Turner2001}.

Grain-surface processes could also form C$_2$H$_3$CHO and C$_2$H$_3$OH in \mbox{TMC-1}. Experiments show that C$_2$H$_3$OH is formed upon proton irradiation of H$_2$O/C$_2$H$_2$ ices \citep{Hudson2003}, and electron irradiation of CO/CH$_4$ and H$_2$O/CH$_4$ ices \citep{Abplanalp2016,Bergantini2017}, while C$_2$H$_3$CHO is produced after electron irradiation of CO/C$_2$H$_4$ ices \citep{Abplanalp2015}. Non-energetic processing of C$_2$H$_2$ ices, in which reactions with H atoms and OH radicals occur on the surface, also produces C$_2$H$_3$OH \citep{Chuang2020}. It remains however uncertain whether these experimental setups (e.g., in terms of irradiation fluxes and ice composition) resemble those of cold dark clouds. \cite{Abplanalp2016} made an effort in this sense by incorporating the results of electron irradiation experiments in a chemical model of a cold dark cloud, and found that cosmic rays could drive the formation of C$_2$H$_3$OH on grain surfaces. Recently, \cite{Shingledecker2019} proposed that C$_2$H$_3$CHO can be efficiently formed on grain surfaces by successive reactions of addition of an H atom to HC$_3$O. This process would also produce propynal (HCCCHO), which make these authors to propose a chemical connection, and thus a potential correlation, between HCCCHO and C$_2$H$_3$CHO. If this mechanism is correct, then C$_2$H$_3$CHO would be more likely detected in those sources with intense HCCCHO emission \citep{Loison2016}.

There is no yet consensus on how are HCOOCH$_3$ and CH$_3$OCH$_3$ formed in cold sources. Models where the synthesis relies on chemical desorption and gas-phase radiative associations usually require a chemical desorption efficiency as high as 10\,\% \citep{Vasyunin2013,Balucani2015,Chang2016}, which can be relaxed if Eley-Rideal processes \citep{Ruaud2015}, radiation chemistry \citep{Shingledecker2018}, or nondiffussive grain-surface processes \citep{Jin2020} are considered. These models can account for abundances relative to H$_2$ around 10$^{-10}$ for HCOOCH$_3$ and/or CH$_3$OCH$_3$ under certain assumptions, although they rely on yet poorly constrained chemical and physical processes. Astronomical observations pointed out that HCOOCH$_3$ and CH$_3$OCH$_3$ could have a chemical connection with CH$_3$OH, based on the slight abundance enhancement inferred for these O-bearing COMs at the CH$_3$OH peak with respect to the dust peak in the pre-stellar core L1544 \citep{Jimenez-Serra2016}. The column densities derived here for HCOOCH$_3$ and CH$_3$OCH$_3$ at \mbox{TMC-1(CP)} are similar, within a factor of two, to those reported by \cite{Soma2018} at the CH$_3$OH peak of \mbox{TMC-1}. A coherent study using the same telescope and a detailed radiative transfer model is needed to see if there is a significant abundance enhancement of HCOOCH$_3$ and CH$_3$OCH$_3$ at the CH$_3$OH peak of \mbox{TMC-1}.

\section{Conclusions}

We reported the detection of C$_2$H$_3$CHO, C$_2$H$_3$OH, HCOOCH$_3$, and CH$_3$OCH$_3$ toward the cyanopolyyne peak of \mbox{TMC-1}. This region, which is a prototypical cold dark cloud with abundant carbon chains, has been revealed as a new cold source where the O-bearing COMs HCOOCH$_3$ and CH$_3$OCH$_3$ are present. In addition, we provide the first evidence of two other O-bearing COMs, C$_2$H$_3$CHO and C$_2$H$_3$OH, in a cold source, the latter being identified unambiguously for the first time in space here. The abundances relative to H$_2$ derived are a few 10$^{-11}$ for C$_2$H$_3$CHO and a few 10$^{-10}$ for the three other molecules. Interestingly, C$_2$H$_3$OH has a similar abundance to its isomer CH$_3$CHO, with C$_2$H$_3$OH/CH$_3$CHO\,$\sim$\,1. We discuss potential formation routes to these molecules and conclude that further experimental, theoretical, and astronomical studies are needed to shed light on the origin of these COMs in cold interstellar sources.

\begin{acknowledgements}

We acknowledge funding support from Spanish MICIU through grants AYA2016-75066-C2-1-P, PID2019-106110GB-I00, PID2019-106235GB-I00, and PID2019-107115GB-C21 and from the European Research Council (ERC Grant 610256: NANOCOSMOS). M.A. also acknowledges funding support from the Ram\'on y Cajal programme of Spanish MICIU (grant RyC-2014-16277). We thank the anonymous referee for a constructive report that helped to improve this manuscript.

\end{acknowledgements}

\end{document}